\documentclass[submission, Proceedings]{SciPost}

\hypersetup{
    colorlinks,
    linkcolor={red!50!black},
    citecolor={blue!50!black},
    urlcolor={blue!80!black}
}

\usepackage[bitstream-charter]{mathdesign}
\DeclareSymbolFont{usualmathcal}{OMS}{cmsy}{m}{n}
\DeclareSymbolFontAlphabet{\mathcal}{usualmathcal}

\begin{document}

\begin{center}{\Large \textbf{
       Future Physics Prospects with the CMS Detector at the High-Luminosity LHC
}}\end{center}

\begin{center}
  C. Henderson\textsuperscript{1} on behalf of the CMS Collaboration
\end{center}

\begin{center}
  {\bf 1} University of Alabama, Tuscaloosa, AL, USA\\
* conor.henderson@ua.edu 
\end{center}

\begin{center}
\today
\end{center}


\definecolor{palegray}{gray}{0.95}
\begin{center}
\colorbox{palegray}{
  \begin{tabular}{rr}
  \begin{minipage}{0.1\textwidth}
    \includegraphics[width=22mm]{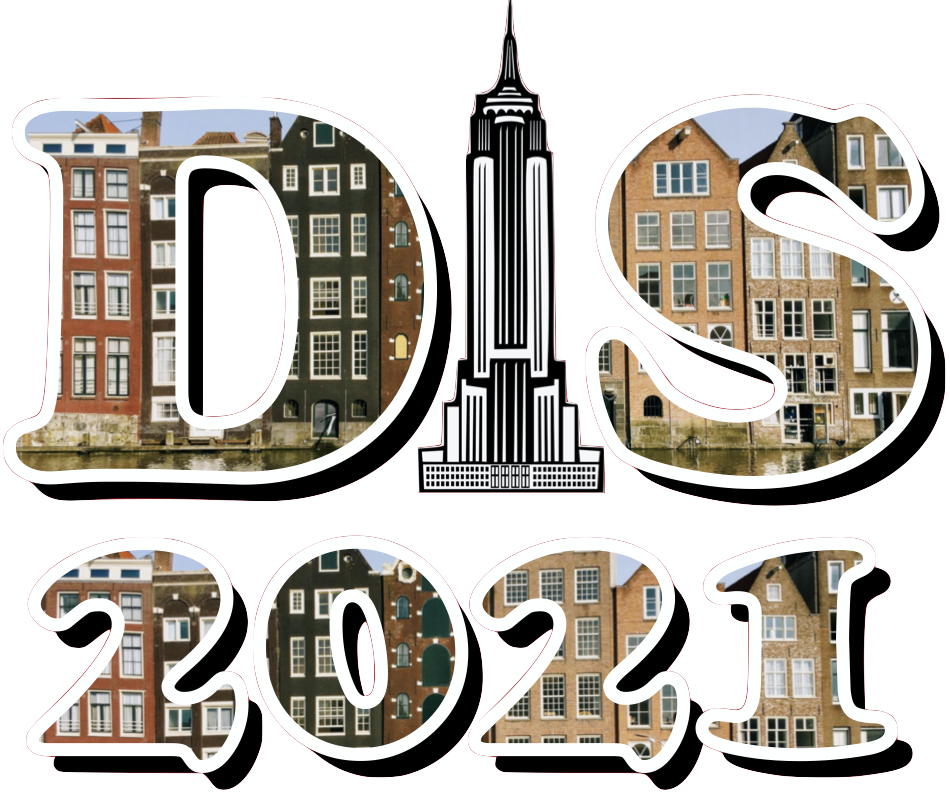}
  \end{minipage}
  &
  \begin{minipage}{0.75\textwidth}
    \begin{center}
    {\it Proceedings for the XXVIII International Workshop\\ on Deep-Inelastic Scattering and
Related Subjects,}\\
    {\it Stony Brook University, New York, USA, 12-16 April 2021} \\
    \doi{10.21468/SciPostPhysProc.?}\\
    \end{center}
  \end{minipage}
\end{tabular}
}
\end{center}

\section*{Abstract}
{\bf
  The High-Luminosity Large Hadron Collider is expected to
  deliver up to 3000~fb$^{-1}$ of proton-proton collisions at 14 TeV
  center-of-mass energy.
  We present prospects for selected
  heavy-ion, Standard Model and Higgs sector measurements with the CMS detector at the HL-LHC, and discuss potential sensitivity to several beyond-Standard Model new physics scenarios.
}


\section{Introduction}
\label{sec:intro}

The LHC plans a future `High-Luminosity' (HL-LHC) phase,
with proton-proton collisions at a center-of-mass energy of 14~TeV
and an instantaneous luminosity  projected to reach a peak of up to
$7.5\times10^{34}$~cm$^{-2}$s$^{-1}$, a factor of roughly four
increase beyond Run 2. The HL-LHC goal is to collect an integrated
luminosity of at least $3000$~fb$^{-1}$ in ten years of
operations~\cite{Apollinari:2015bam}.
The CMS experiment~\cite{Chatrchyan:2008aa} is preparing for a set of major detector upgrades
for the HL-LHC~\cite{Contardo:2020886}.
In these proceedings, we present prospects for selected
  heavy-ion, Standard Model (SM) and Higgs sector measurements with the CMS
  detector at the HL-LHC, and discuss potential sensitivity to several
  beyond-Standard Model (BSM) new physics scenarios.
  The complete set of public CMS physics projections for
 the HL-LHC are available at Ref~\cite{cms_pubs_page_futurephysics}.

    \section{Heavy-Ion Physics Projections for the HL-LHC}

  
  
\begin{figure}[htb]
    \begin{center}
      \includegraphics[trim= 0mm 3mm 0mm 0mm ,clip, width=0.45\textwidth]{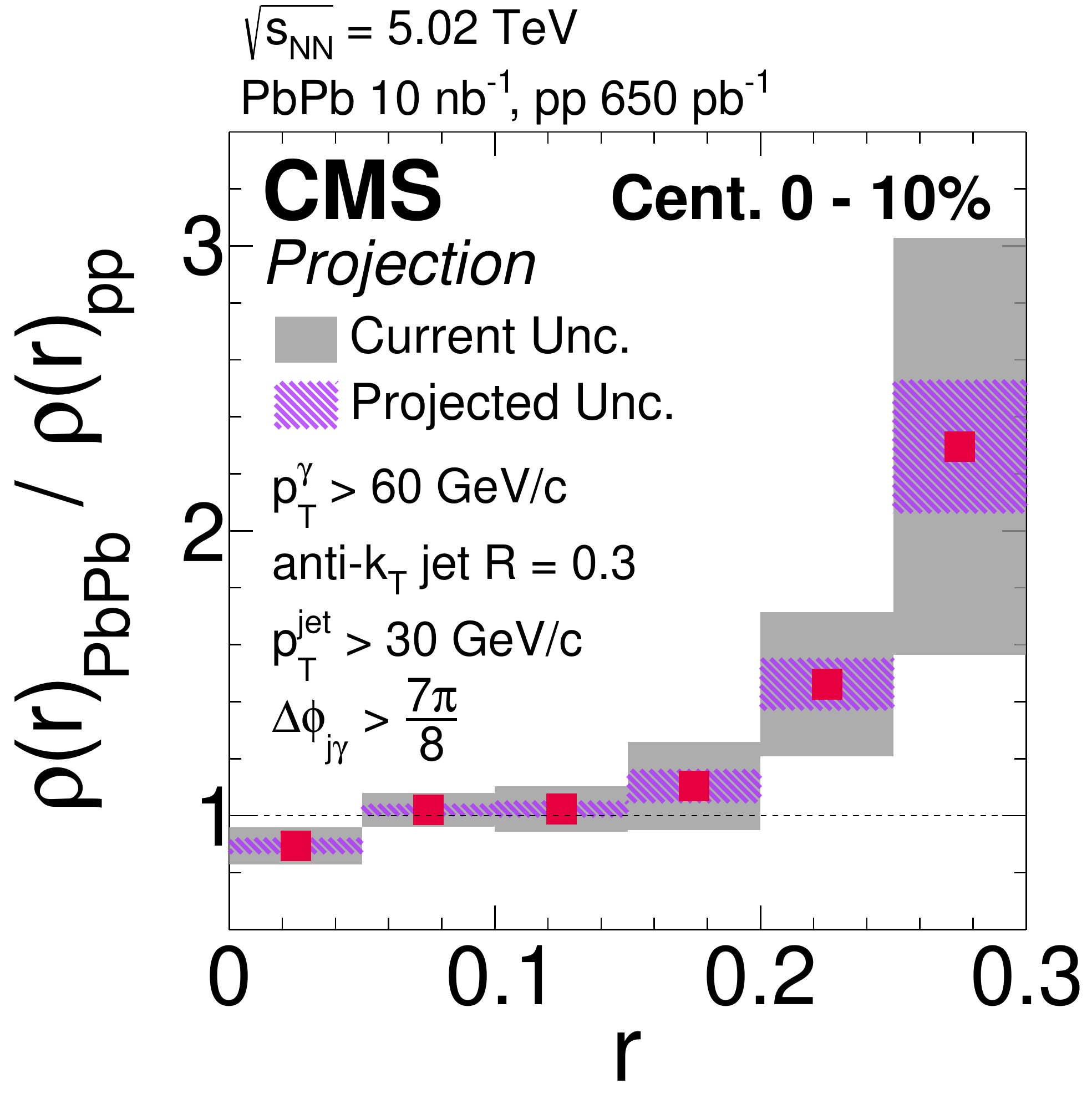}
    \end{center}
    \caption{Ratio of the density of particles as a function of the
      radius $r$ from the central axis of a photon-tagged jet in
      central PbPB collisions relative to $pp$ collisions~\cite{CMS-PAS-FTR-18-025}. Projected
      systematic uncertainties on the HL-LHC measurement are shown
      compared to current uncertainties. }
    \label{fig:hi_jetquench}
\end{figure}

The suppression or modification of high-energy jets observed in
heavy-ion collisions, known as jet quenching, is interpreted as being
caused by strong interactions between the high-energy parton and the
deconfined colored medium created in the heavy-ion collision.
Heavy-ion collisions at the HL-LHC will allow for more detailed
measurements of this phenomenon.
Figure~\ref{fig:hi_jetquench} shows the projection by CMS for a
measurement at the HL-LHC of the ratio of the density of particles
produced at a radius $r$ from the central axis of a photon-tagged jet
in central PbPb collisions, relative to $pp$
collisions~\cite{CMS-PAS-FTR-18-025}.
The projected significant reduction in systematic uncertainties on
this measurement at the HL-LHC will allow for more detailed probes of the underlying
parton-medium interaction.

Heavy-ion collisions at the HL-LHC can also constrain more precisely the
parton density functions (PDFs) within the nucleus. For instance, the
photoproduction of $\Upsilon(1S)$ is proportional to the gluon
density, and therefore a measurement can constrain the gluon
shadowing factor (the ratio of gluon density in a nucleus compared to
a proton). A CMS projection for PbPb collisions at
the HL-LHC~\cite{CMS-PAS-FTR-18-027} shows that this measurement could
constrain the gluon shadowing factor down to a Bjorken-$x$ value
around $10^{-4}$, which would be a significant extension of the
current $x$ range probed. 


  \section{Higgs Sector Projected Results for the HL-LHC}

\begin{figure}[htb]
    \begin{center}
      \includegraphics[trim= 0mm 0mm 0mm 0mm ,clip,
      width=0.45\textwidth]{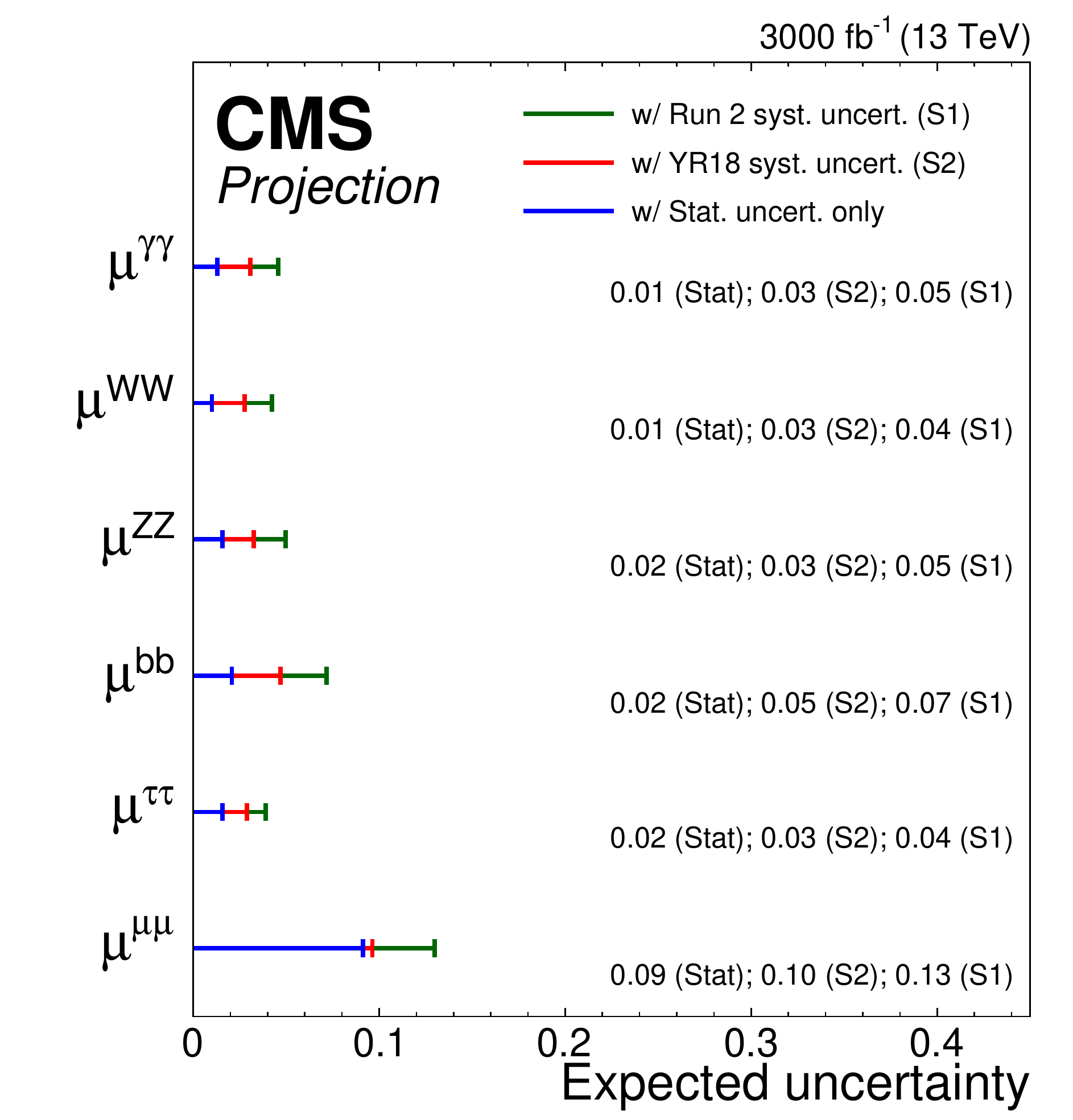}
        \includegraphics[trim= 0mm 0mm 0mm 0mm ,clip, width=0.45\textwidth]{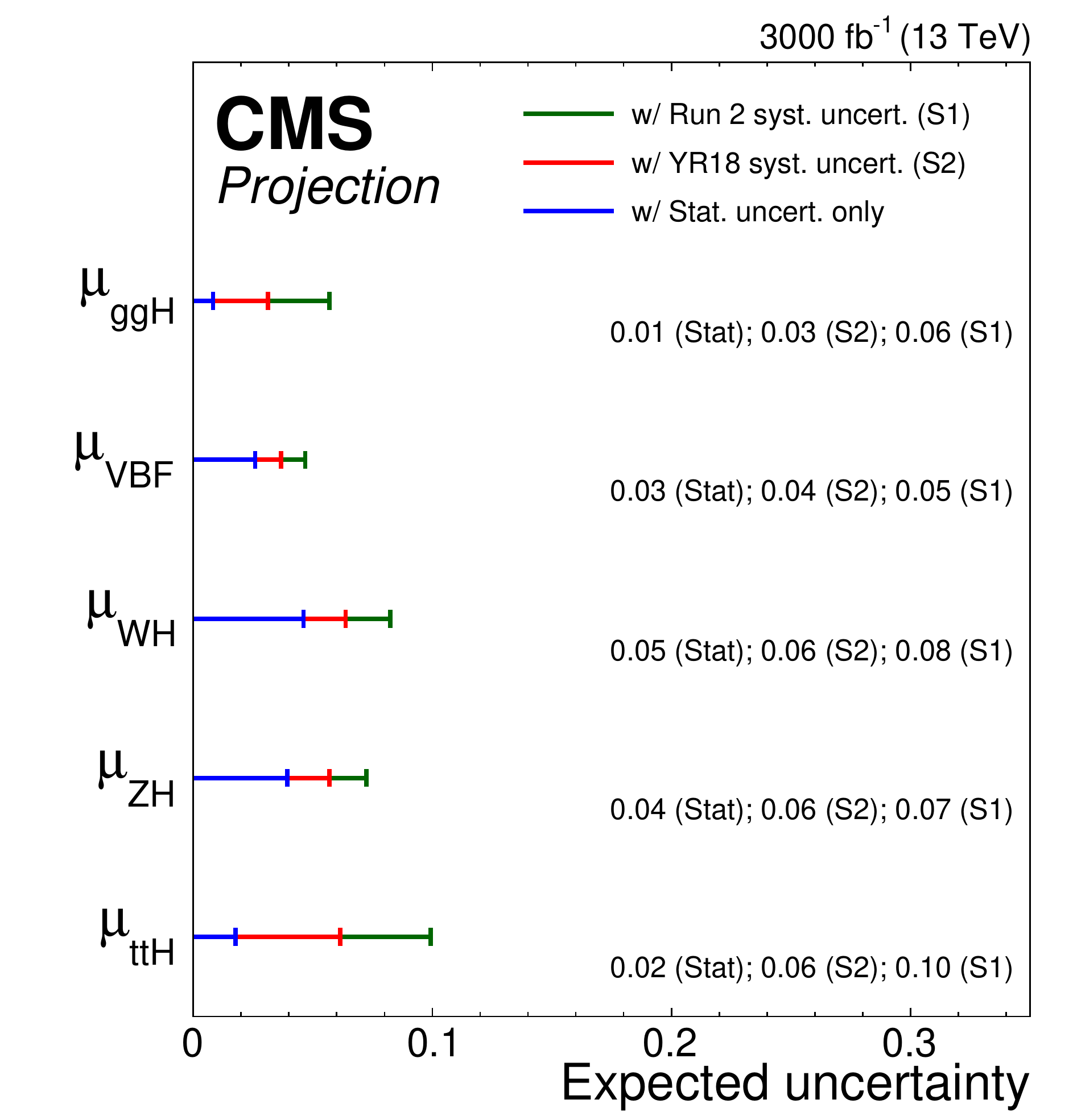}
    \end{center}
    \caption{Projected CMS uncertainties from the HL-LHC on the Higgs coupling strength
      parameters for various channels (left) and production mechanisms
    (right)~\cite{CMS-PAS-FTR-18-011}.}
    \label{fig:higgs_couplings}
\end{figure}

A crucial test for our understanding of the Higgs sector is to measure
the Higgs boson coupling strengths to other SM particles, which are all
constrained precisely in the SM, and therefore any observed
deviations from the predicted values would represent evidence for
beyond-SM physics.
CMS has performed a projection of the precision expected for
these measurements with 3000~fb$^{-1}$ at
the HL-LHC~\cite{CMS-PAS-FTR-18-011}, and the results are shown in
Figure~\ref{fig:higgs_couplings}, in terms of $\mu$, the coupling strength
parameter per channel.
Overall, the total uncertainty is projected to be around
5\% for most channels, with a larger value of $\sim 10\%$
for the rarer $H \rightarrow \mu \mu$ channel; thus these HL-LHC
measurements should be a signficant test of the SM predictions for the Higgs channel couplings.   
A comparable level of experimental precision is projected for
measurements of the Higgs boson production mechanisms 
as shown in the second plot of the same figure. 
  
 
In the SM, the Higgs potential is  $V(\phi) = -\mu^2
\phi^\dagger\phi + \lambda (\phi^\dagger\phi)^2$, where $\lambda$
represents the Higgs field self-coupling, and the SM parameter values are
determined once the Higgs boson mass is known.
The Higgs field self-coupling contributes to di-Higgs production --
observation of this rare process can therefore probe the self-coupling
parameter and would act as an important independent test of the actual
Higgs potential and hence the overall Higgs sector. 
CMS has performed a projection for its ability to measure di-Higgs
production at the HL-LHC~\cite{CMS-PAS-FTR-18-019}.
Five HH decay channels have been considered ($HH \rightarrow
bbbb,bb\tau\tau, bbWW, bbZZ$ and $bb\gamma\gamma$), and the combination
is projected to yield 2.6 sigma significance for SM HH production. 
In terms of the coupling modifier $\kappa_\lambda$
(the ratio of the parameter relative to the SM expectation),
it is projected that, at 95\% CL, $\kappa_\lambda$ could be
constrained within the range $[-0.18,3.6]$.


  \section{Projected Top Quark Measurements at the HL-LHC}

\begin{figure}[htb]
    \begin{center}
      \includegraphics[trim= 0mm 0mm 0mm 0mm ,clip,
      width=0.4\textwidth]{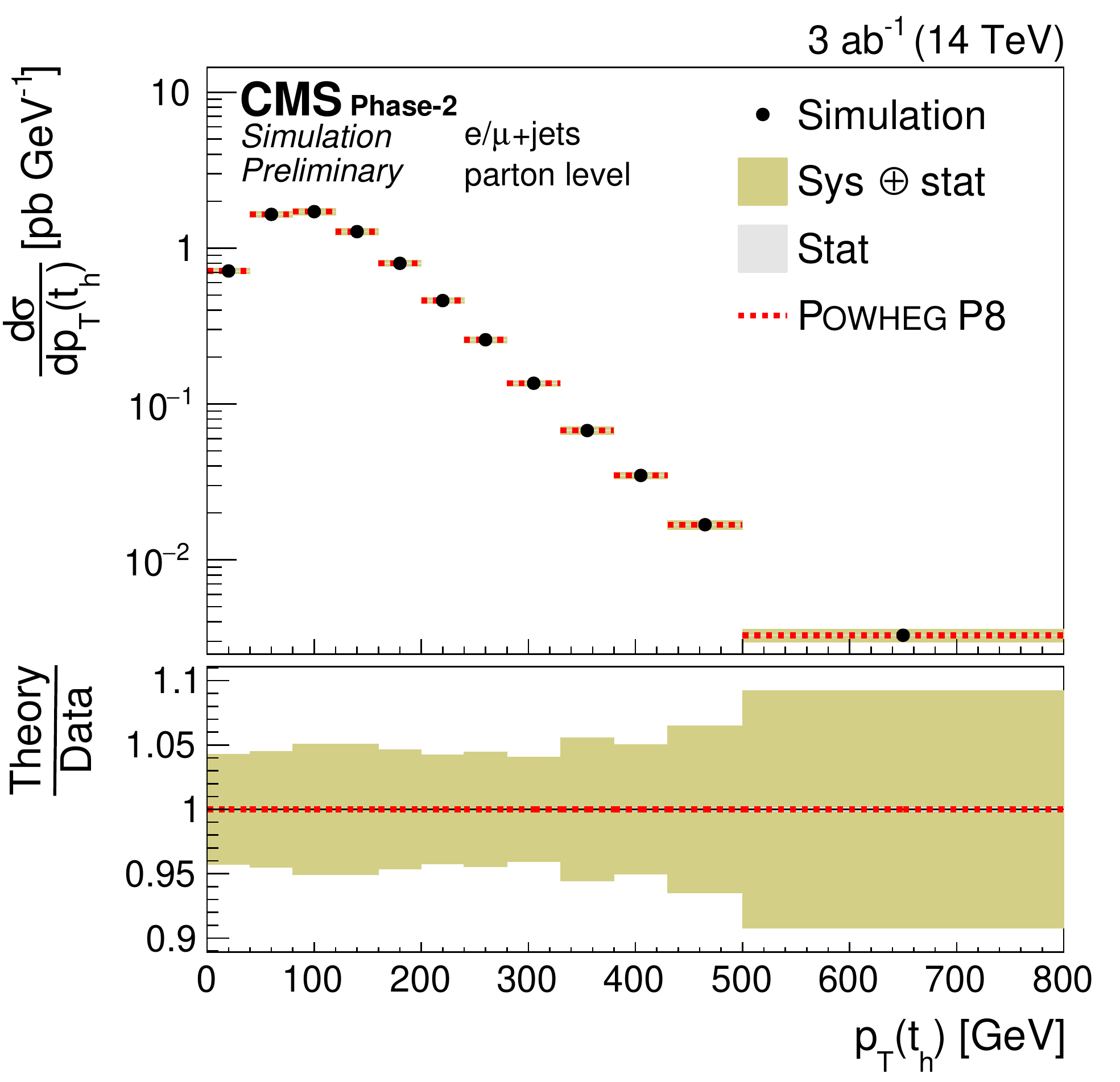}
      \includegraphics[trim= 0mm 0mm 0mm 0mm ,clip,
      width=0.4\textwidth]{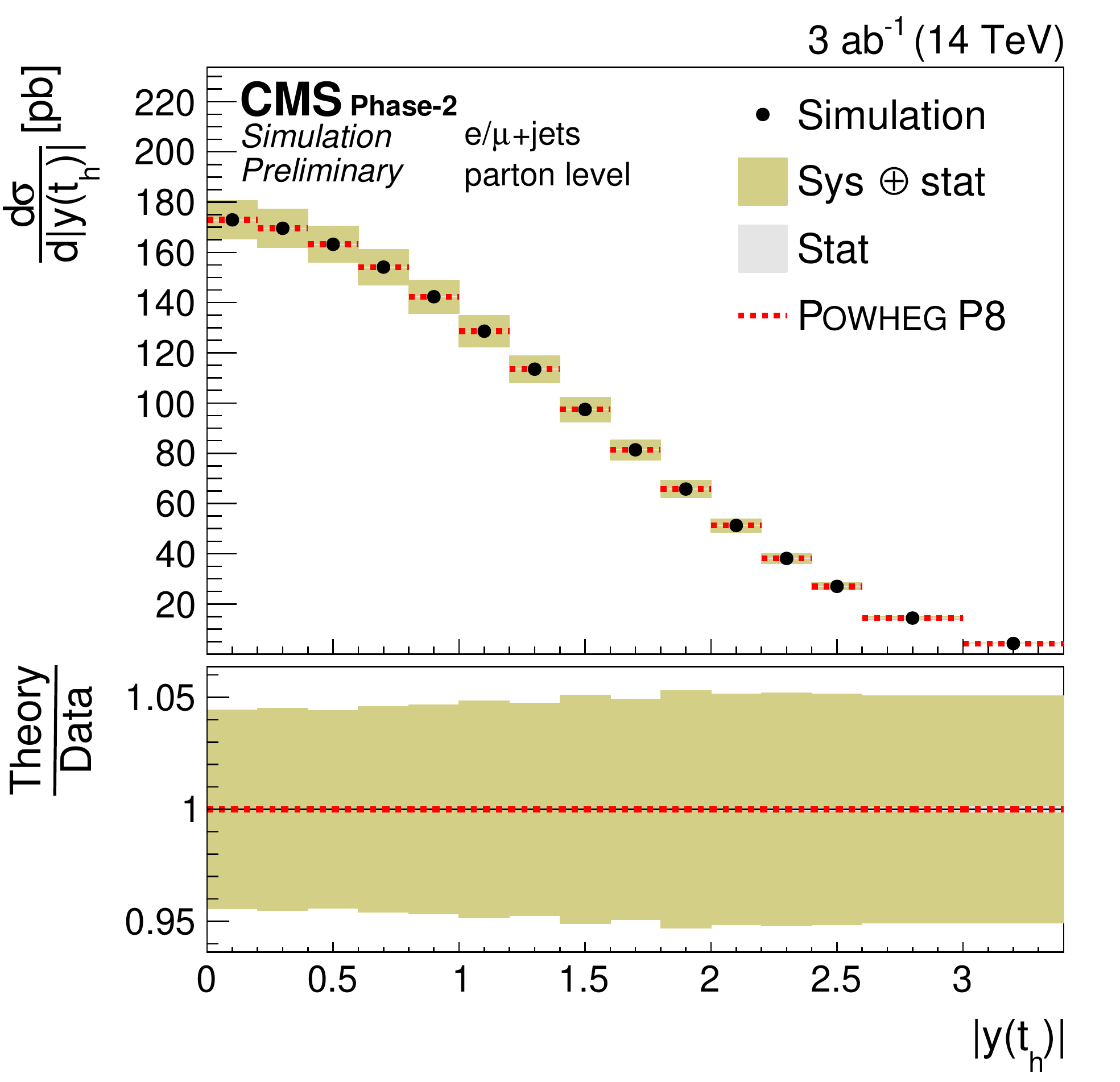}
    \end{center}
    \caption{Projected $t\bar{t}$ differential cross sections as a
      function of the top quark transverse momentum (left) and
      rapidity (right)~\cite{CMS-PAS-FTR-18-015}.}
    \label{fig:ttbar_diff_xsecs}
\end{figure}

An illustration of the potential capability of the HL-LHC for precision
physics in the top quark sector comes is the CMS projection
for $t\bar{t}$  differential cross section measurements~\cite{CMS-PAS-FTR-18-015}.
Figure~\ref{fig:ttbar_diff_xsecs} shows the projected results as a
function of the top quark $p_T$ and rapidity. 
The systematic uncertainties on this measurement are projected to be
in the range 5-10\%, representing about a factor of two
improvement over the equivalent LHC Run 2 results, and will allow
for a precision test of the SM top quark theoretical predictions at
the HL-LHC. 

The large luminosity obtained from the HL-LHC will also allow probes of
even rarer top quark processes, such as the production of four top
quarks, where observed differences relative to the SM expectation for
this rare process could indicate contributions from BSM physics.
For the HL-LHC, CMS projects that the cross section for
SM $t\bar{t}t\bar{t}$ production could be constrained to the level of
18-28\%~\cite{CMS-PAS-FTR-18-031}.

  \section{Searches for Dark Matter and Heavy Resonances at the HL-LHC}

\begin{figure}[htb]
    \begin{center}
      \includegraphics[trim= 0mm 5mm 0mm 0mm ,clip,
      width=0.45\textwidth]{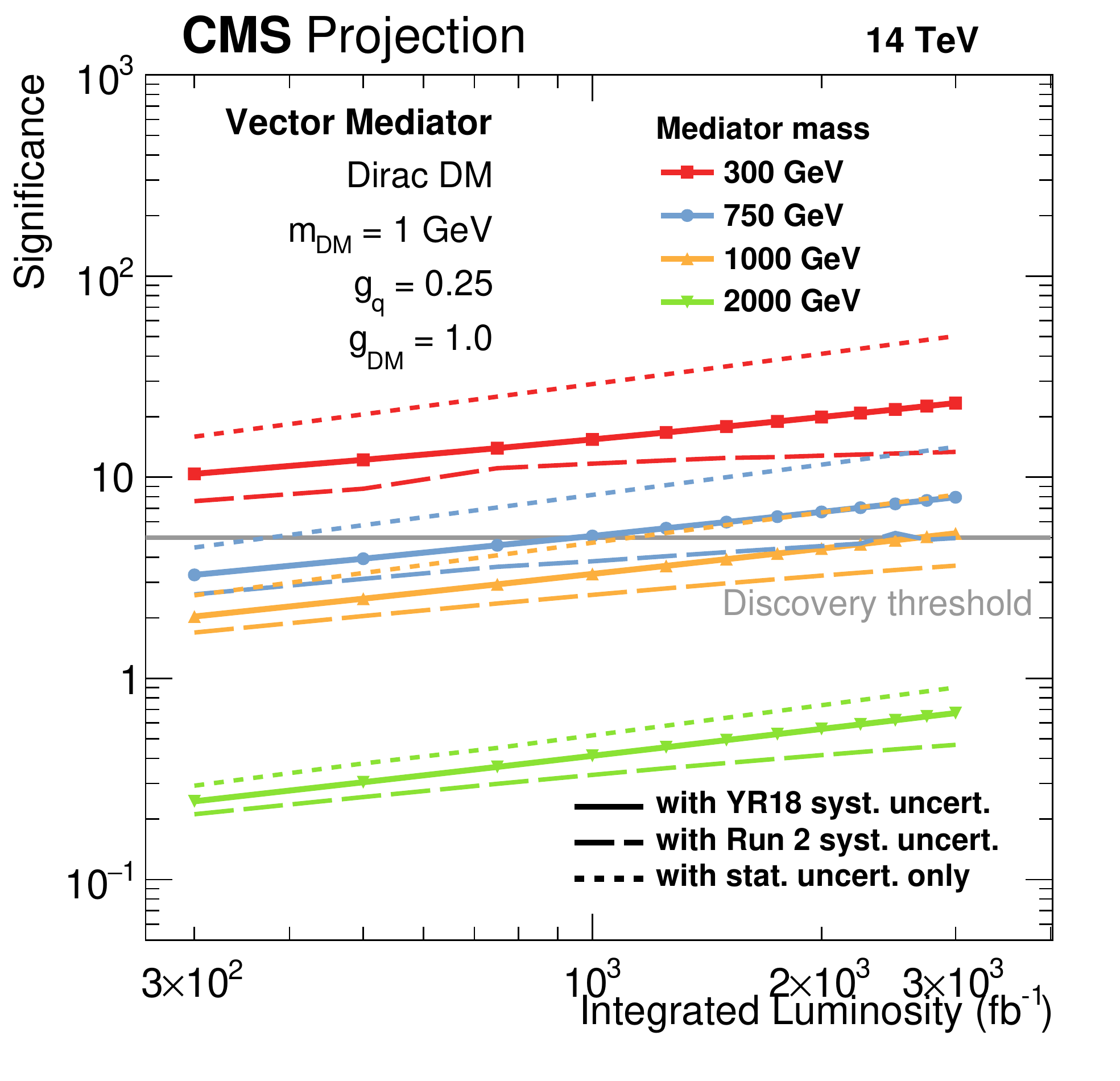}
    \end{center}
    \caption{Discovery potential as a function of the HL-LHC
      integrated luminosity, for Dark Matter vector mediators of
      various masses in the Z boson +missing   energy search channel~\cite{CMS-PAS-FTR-18-007}. }
    \label{fig: monoZ_dm_limits}
\end{figure}

  
  One way to search for Dark Matter (DM) at the LHC is to look for
  new invisible particles in events
  containing a Z boson and missing transverse momentum.
  CMS has performed a projection for this analysis at
  the HL-LHC~\cite{CMS-PAS-FTR-18-007}. A benchmark scenario is considered
  with a vector mediator for Dirac DM, assuming the DM
  particle has mass 1 GeV and the SM/DM coupling parameters are fixed to
  certain values. 
  Figure~\ref{fig: monoZ_dm_limits} shows the projected discovery
  potential
  as a function of the integrated luminosity
  at the HL-LHC; it can be seen that for 3000~fb$^{-1}$, 5$\sigma$
  discovery is possible for DM vector mediators with mass up to about 1~TeV.

Projections have also been performed for new heavy resonances decaying
to   ZZ~\cite{CMS-PAS-FTR-18-040} and
$t\bar{t}$~\cite{CMS-PAS-FTR-18-009}. 
  In the former, the final state considered is $ZZ\rightarrow2\ell
  2q$, including both resolved and merged-jet categories. Assuming a
  scalar resonance with a natural width much smaller than the
  experimental  resolution, the analysis projects to be able to
  exclude at 95\% CL cross sections for this signal process of
  approximately $0.1-2$~fb, in the $M_{ZZ}$ range $1-3$~TeV.
  For a $t\bar{t}$ resonance,
  the decay of an excited state of a gluon in a Randall-Sundrum extra
  dimension model is considered as a benchmark scenario.
  The analysis projects to have discovery potential
  with $5\sigma$ significance for masses of
  the Randall-Sundrum gluon up to 5.7~TeV

\section{Conclusion}
In conclusion, the High-Luminosity Large Hadron Collider aims to
collect an integrated luminosity of $3000$~fb$^{-1}$ in ten years of
operations, at a collision energy of 14~TeV.
We have presented here a selection of projections by the CMS experiment that
illustrate the broad physics capabilities from the HL-LHC, across heavy-ion
physics, the Higgs sector, top quark physics and beyond-Standard Model
searches. 
 The complete set of public CMS physics projections for
  the HL-LHC are available at Ref~\cite{cms_pubs_page_futurephysics}. 

\section*{Acknowledgements}
The author is supported by the DOE under award DE-SC0012447.

\bibliography{henderson_dis2021_proceedings_futurephysics.bib}

\nolinenumbers

\end{document}